\documentclass[journal]{IEEEtran}

\usepackage{graphicx}
\usepackage{booktabs,tabularx,makecell,adjustbox}
\usepackage{array}
\usepackage{amsmath}
\usepackage{url}
\usepackage{cite}
\usepackage{microtype}
\graphicspath{{figures/}}
\makeatletter
\renewcommand{\@maketitle}{%
  \newpage
  \begingroup
  \centering\normalfont
  \vskip 10pt
  {\fontsize{18}{22}\selectfont\bfseries\@title\par}
  \vskip 24pt
  {\normalsize\@author\par}
  \endgroup
}
\makeatother

\begin{document}

\title{Wav2Aug: A Drop-In Speech Augmentation Policy}

\author{\textbf{Gianfranco Dumoulin Bertucci}\textsuperscript{1,2,*}\qquad
\textbf{Mirco Ravanelli}\textsuperscript{1,2}\\[0.3em]
\textsuperscript{1}Concordia University\quad
\textsuperscript{2}Mila-Quebec AI Institute\\[0.2em]
{\small\texttt{gianfranco.dumoulin@mail.concordia.ca}\quad\texttt{mirco.ravanelli@concordia.ca}}
\thanks{\textsuperscript{*}Corresponding author.}
}

\IEEEaftertitletext{\vspace{-2\baselineskip}}
\maketitle

\begin{abstract}
SpecAugment is the dominant speech augmentation policy, but researchers have studied it primarily for automatic speech recognition (ASR), and it operates on spectrograms, limiting its applicability to other speech tasks and waveform models. We propose Wav2Aug, a drop-in waveform augmentation policy that randomly samples two augmentations from a calibrated pool of waveform transformations. We calibrate the augmentation hyperparameters once and reuse them across tasks and architectures, eliminating task-specific tuning. Experiments on ten speech benchmarks show that Wav2Aug achieves ASR performance comparable to SpecAugment while improving over it on average across non-ASR tasks, yielding an average relative error reduction of 12.8\% over the strongest compatible SpecAugment baseline. We further show that multi-view training provides additional gains, with two augmented views offering the best tradeoff between performance and computational cost.
\end{abstract}


\IEEEpeerreviewmaketitle

\section{Introduction}

Data augmentation improves speech model robustness by increasing training diversity without collecting new data \cite{dataaug}. Speech augmentation policies often combine perturbations, masking operations, and randomized transformation pools \cite{contaminated_speech_mirco,speed_noise,Ravanelli2016Realistic,sign_flip,amplitude,impulse_mirco, babble,realm, specaug,randaug,audrandaug}. 
A prominent example is SpecAugment \cite{specaug}, which applies masking and warping operations in the time and frequency domain.
Although more sophisticated policy search and adaptive augmentation methods have been proposed \cite{gaugment,autoaugment,sapaugment}, SpecAugment remains the dominant augmentation strategy in speech processing. It has even been used in large-scale systems such as Whisper \cite{whisper} and Qwen-Audio \cite{qwen}, despite some important limitations.

SpecAugment has been extensively validated on ASR tasks, but its effectiveness on other speech processing tasks remains less studied. We therefore evaluate SpecAugment as a drop-in augmentation policy for non-ASR speech tasks using its original fixed configurations \cite{specaug, wav2vec2} as done in \cite{whisper,qwen}, and observe substantially smaller gains than those typically reported for ASR.

Moreover, SpecAugment transfers less well to models that do not use spectrogram inputs. Self-supervised speech encoders such as wav2vec~2.0 \cite{wav2vec2}, HuBERT \cite{hubert}, and WavLM \cite{wavlm}, for example, cannot use standard SpecAugment directly and instead apply similar masking operations to latent features.

Beyond these limitations, many studies, including \cite{lu2020sentiment,le2021adapters,lee2022s2st,tang2021auxiliary,tang2023hybrid,ma2024simultaneous,rouditchenko2024flamingo}, adopt standard SpecAugment configurations without reporting task-specific tuning of the augmentation hyperparameters. We hypothesize that the widespread use of these "frozen" policies is due to the computational cost of tuning augmentation hyperparameters and interpret the trend as evidence of a practical need for drop-in augmentation.

Motivated by these gaps, we introduce Wav2Aug, a fixed waveform augmentation policy applied on the fly to raw audio before feature extraction. 
Wav2Aug samples two augmentations per batch from a large pool of carefully selected transformations and applies them sequentially to each waveform. Unlike AudRandAug \cite{audrandaug}, which tunes the number of operations and their magnitude, Wav2Aug selects hyperparameter ranges once and then reuses them unchanged across all datasets and architectures, enabling task-agnostic drop-in use.

\begin{figure}[t]
  \centering
  \includegraphics[width=1\linewidth]{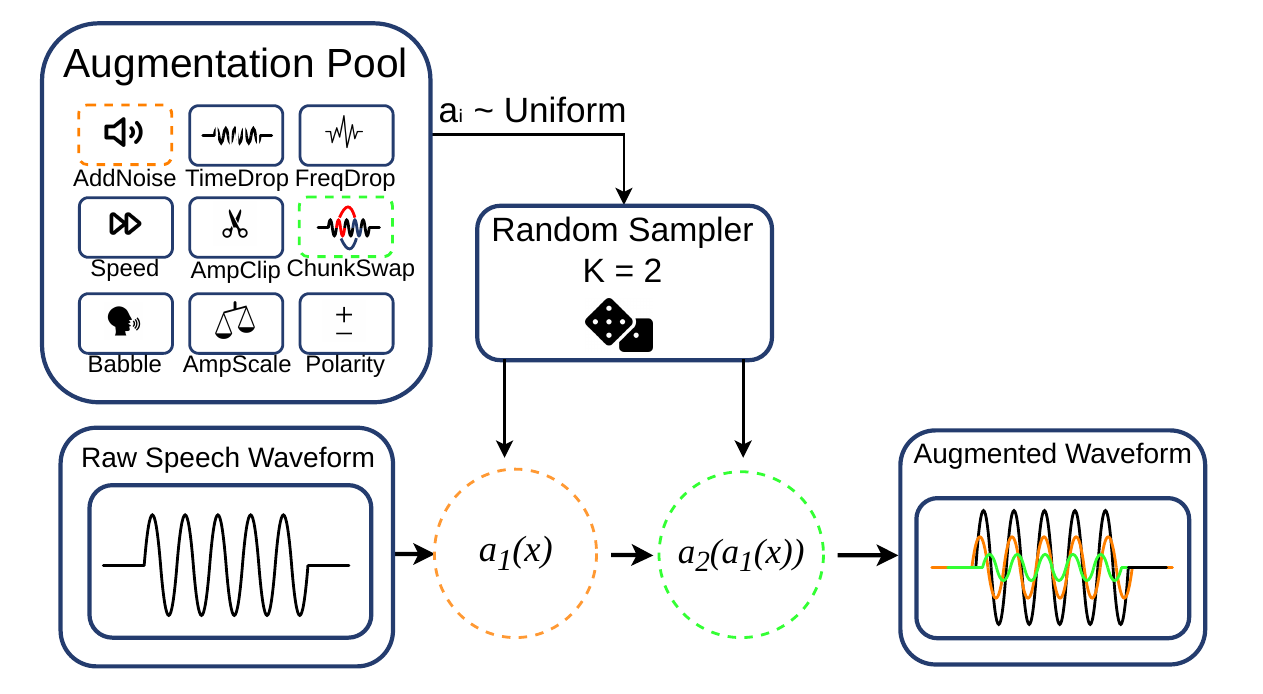}
  \caption{Overview of Wav2Aug. We sample two augmentations uniformly per batch and apply them sequentially to each utterance, with hyperparameters drawn from globally fixed ranges.}
  \label{fig:wav2aug}
\end{figure}

We benchmark Wav2Aug primarily as a drop-in policy for non-ASR speech tasks. Wav2Aug achieves lower mean error than the strongest compatible SpecAugment baseline on all five evaluated non-ASR tasks, with an average relative error reduction of 12.8\%. We also evaluate ASR performance on speech benchmarks and show that Wav2Aug remains competitive with SpecAugment.

We also study multi-view augmentation, where we concatenate multiple independently augmented versions of the same utterance within a batch. Multiple views expose the model to several independently sampled perturbations within the same update \cite{hoffer2020augment,fort2021drawing,kivaisi2024minibatch,singh2024roar}. We find that the best-performing Wav2Aug configuration uses two concatenated augmented views, improving performance by 19.6\% on average relative to non-augmented training.

Like SpecAugment, Wav2Aug offers practical defaults that may not be optimal for every task but perform well across a wide range of speech processing tasks without any augmentation hyperparameter tuning.

We release the parameter ranges, code, and an installable Python library as open source\footnote{\url{https://github.com/gfdb/wav2aug}}.

\section{Wav2Aug}

\begin{figure}[t!]
  \centering
  \includegraphics[width=\columnwidth]{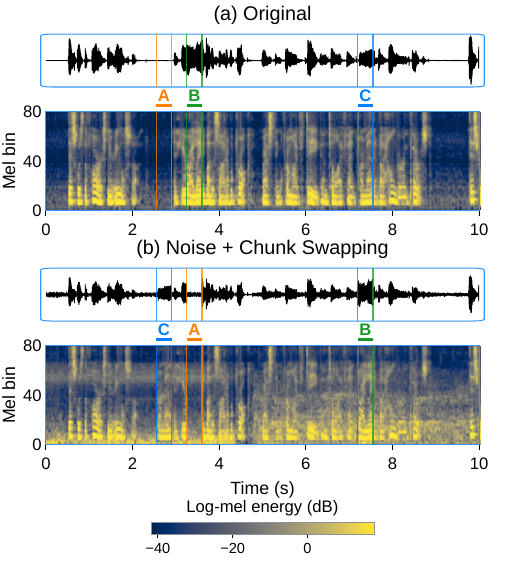}
  \caption{Waveform and log-mel spectrogram before and after noise addition followed by chunk swapping. Colored lines mark chunk boundaries. Matching letters track swapped chunks.}
  \label{fig:example}
\end{figure}

Figure~\ref{fig:wav2aug} shows the proposed augmentation policy, which randomly samples two distinct augmentations from the pool and applies them sequentially to all waveforms in a batch.

\subsection{Augmentation Pool} 
Each operation targets a different kind of variability in speech recordings.
Figure~\ref{fig:example} shows an example waveform and spectrogram before and after applying Wav2Aug.
The pool includes waveform versions of SpecAugment-style masking operations, together with several additional transformations.

\begin{table}[t]
\centering
\small
\setlength{\tabcolsep}{4pt}
\renewcommand{\arraystretch}{1.05}
\caption{Wav2Aug augmentations and fixed ranges at 16 kHz. Rows are ordered by individual calibration effectiveness, the error reduction obtained when each operation is evaluated alone.}
\label{tab:waveform-augs}
\begin{tabular*}{\columnwidth}{@{\extracolsep{\fill}}l l l r@{}}
\toprule
\textbf{Rank} & \textbf{Augmentation} & \textbf{Hyperparameter} & \textbf{Range / Value} \\
\midrule
1 & Noise Add. & SNR (dB) & $[0,10]$ \\
\addlinespace[0.4ex]
2 & Freq. Dropout & Drop bounds & $[0,1]$ \\
  &               & Band count  & $[1,8]$ \\
  &               & Band width  & $0.1$ \\
\addlinespace[0.4ex]
3 & Time Dropout & Chunk count       & $[1,8]$ \\
  &              & Chunk length (ms) & $[0,250]$ \\
\addlinespace[0.4ex]
4 & Speed Pert. & Speed factor & $\{0.9,1.0,1.1\}$ \\
\addlinespace[0.4ex]
5 & Amp. Clip. & Clip threshold & $[0,0.75]$ \\
\addlinespace[0.4ex]
6 & Chunk Swap. & Num. chunks     & $4$ \\
  &             & Chunk size (\% Len.) & $0.01$ \\
\addlinespace[0.4ex]
7 & Babble Add. & SNR (dB) & $[0,20]$ \\
\addlinespace[0.4ex]
8 & Amp. Scale & Gain & $[0.05,0.5]$ \\
\addlinespace[0.4ex]
9 & Polarity Inv. & Probability & $0.6$ \\
\bottomrule
\end{tabular*}
\end{table}

\textbf{Noise addition} randomly selects a noise segment from an available noise dataset. We use the point-source noise subset of the OpenRIR corpus \cite{pointsource}, chosen for its broad adoption and open-source availability. We crop or pad the selected segment to match the utterance length and mix it with the input waveform at a randomly sampled signal-to-noise ratio (\textit{SNR}). This augmentation improves robustness to environmental background noise while operating directly on the raw waveform.

\textbf{Frequency dropout} removes randomly selected narrow frequency bands directly in the waveform domain using notch filters. We randomly sample the frequency range (\textit{drop bounds}), number of bands (\textit{band count}), and bandwidths (\textit{band widths}) of the dropped bands, encouraging the model to rely on complementary spectral cues while preserving the waveform representation.

\textbf{Time dropout} removes randomly selected temporal regions by zeroing contiguous waveform segments. We independently sample the number (\textit{chunk count}), duration (\textit{chunk length}), and position of the dropped segments for each utterance, improving robustness to local temporal corruptions and missing speech regions.

\textbf{Speed perturbation} changes the speaking rate by resampling the waveform with a randomly selected \textit{speed factor}. This augmentation simulates natural variability across speakers and speaking styles by jointly modifying speech tempo and pitch.

\textbf{Amplitude clipping} simulates microphone or recording saturation by clipping waveform amplitudes above a randomly sampled threshold (\textit{Clip threshold}), improving robustness to nonlinear distortions.

\textbf{Chunk swapping} randomly exchanges waveform segments (\emph{Num. Chunks}), each with a duration specified by \emph{Chunk Size}, perturbing local temporal structure while preserving overall signal content and duration.

\textbf{Babble noise addition} follows the same SNR-based mixing procedure as noise addition, but constructs the interference signal by summing the other utterances in the current minibatch. This simulates overlapping speakers without requiring a separate babble-noise corpus.

\textbf{Amplitude scaling} multiplies each waveform by a randomly sampled \textit{gain} factor, increasing robustness to variations in recording volume.

\textbf{Polarity inversion} flips the waveform's sign with a fixed \textit{probability}, simulating phase reversals introduced by different recording or playback devices.

\subsection{Calibration}
\label{sec:hyperparam_calib}
Each augmentation has hyperparameters and sampling ranges, reported in Table~\ref{tab:waveform-augs}. To enable task-agnostic deployment, we calibrate the ranges and the number of sequential operations, $K$, once. For each augmentation, we define candidate configurations, evaluate each independently, and select the one that performs best across the calibration datasets. We evaluate hundreds of candidate configurations via grid search, varying parameters such as SNR, dropout count, and dropout width.

Formally, let $a$ denote an augmentation, $c$ a candidate configuration for that augmentation, and $d$ a calibration dataset. We write $C_a$ for the set of candidate configurations for augmentation $a$, and $D$ for the set of calibration datasets. Let $E_{d,0}$ be the error without augmentation and $E_{d,c,a}$ the error obtained by applying augmentation $a$ with configuration $c$. The improvement is then defined as
\begin{equation}
\Delta_{d,c,a}=E_{d,0}-E_{d,c,a}.
\end{equation}
Positive values of $\Delta_{d,c,a}$ indicate that the candidate reduces error relative to no augmentation, while negative values indicate that it degrades performance.
For each dataset and augmentation, we compute the mean and standard deviation over all candidates $c \in C_a$
\begin{align}
\mu_{d,a} &= \frac{1}{|C_a|}\sum_{c \in C_a}\Delta_{d,c,a},\\
\sigma_{d,a} &= \sqrt{\frac{1}{|C_a|-1}\sum_{c \in C_a}(\Delta_{d,c,a}-\mu_{d,a})^2}.
\end{align}
Each candidate is then scored by its average normalized gain
\begin{equation}
Z_{c,a}=\frac{1}{|D|}\sum_{d \in D}\frac{\Delta_{d,c,a}-\mu_{d,a}}{\sigma_{d,a}}.
\end{equation}
Among the candidates, we keep the one with the largest $Z_{c,a}$. The same process gives $K=2$ operations per batch.

We then hold the resulting ranges fixed and evaluate them on separate ASR and non-ASR tasks to test whether the policy transfers without task-specific tuning. We use ASR for calibration because Wav2Aug was initially designed to match or improve upon SpecAugment on speech recognition benchmarks. After discovering Wav2Aug could not topple the bastion, but that it nearly matched its performance, we decided to explore the frozen policies' performance on non-ASR.

Calibration uses three ASR settings rather than a single dataset or architecture. They span different languages, model families, training objectives, and acoustic conditions. These settings are LibriSpeech-100h \cite{librispeech} with a Transformer, Common Voice Italian \cite{commonvoice} with wav2vec~2.0 (CTC), and VoiceBank \cite{voicebank} with a CRDNN-Transducer.
For comparison, SpecAugment baselines are evaluated using the same datasets and SpeechBrain recipes. All calibration experiments use the default optimization hyperparameters from those recipes \cite{speechbrain_v1}.

Table~\ref{tab:waveform-augs} summarizes the final augmentation pool and calibrated ranges. The order reflects the individual error reduction each operation achieved during calibration when evaluated alone. Noise addition is strongest, while frequency and time dropout also rank highly, mirroring masking in SpecAugment but operating directly on waveforms. Compared with common SpecAugment masking settings, Wav2Aug uses shorter time drops and multiple narrow frequency drops. Speed perturbation also performs strongly, consistent with prior work \cite{speed_noise}. 

These rankings describe operations evaluated individually; they do not establish which combination works best. We therefore also examine how performance changes with the number of available operations. A top-$k$ pool contains the first $k$ operations in Table~\ref{tab:waveform-augs}, with their calibrated ranges unchanged. We compare two fixed choices across all ten benchmarks. Top-6 adds noise via chunk swapping, and top-9 includes all nine operations. Both sample two operations per batch. We use top-6 by default and keep top-9 as a larger alternative; the pool-size study in Section~\ref{sec:poolsize} examines this choice.

\section{Experiments}

\subsection{Evaluation protocol}
For the ten-benchmark comparison, we evaluate two fixed SpecAugment settings, LibriSpeech Basic (LB) and Switchboard Mild (SM) \cite{specaug}. These operate on time-frequency features. For wav2vec~2.0 and HuBERT, which take waveform inputs, we instead use the time- and channel-masking settings on latent features recommended in the original works, via the Hugging Face implementations \cite{wav2vec2,hubert,huggingface}. Comparisons to the strongest compatible SpecAugment baseline use the lowest mean error among the settings available for each model.

For each dataset and augmentation policy in these experiments, we independently tune the learning rate and weight decay using a $3 \times 3$ grid around the default SpeechBrain recipe values, with AdamW as the optimizer. This lets us evaluate each augmentation policy under optimization settings suited to its training dynamics. This tunes model training, not the augmentation ranges, which remain fixed. Other recipe-specific augmentations are disabled so that the comparison isolates the policy being evaluated. We report means and standard deviations over five random seeds.

\begin{table*}[t]
\centering
\caption{\normalfont ASR test WER (\%), with PER for TIMIT and CER for AISHELL-1. Mean $\pm$ standard deviation over five seeds; lower is better. LB and SM denote the LibriSpeech Basic and Switchboard Mild SpecAugment policies. Parameter counts are approximate.}
\label{tab:asr}
\small
\setlength{\tabcolsep}{4pt}
\renewcommand{\arraystretch}{1.12}
\begin{adjustbox}{max width=\textwidth}
\begin{tabular}{l *{5}{c}}
\toprule
Augmentation Policy
& \makecell{CommonVoice AR\\ASR (54h) \\Whisper \cite{whisper}\\(764M)}
& \makecell{TIMIT\\ASR (5h)\\CRDNN\\(10M)}
& \makecell{AISHELL-1\\ASR ($\sim$170h)\\Transformer\\(31M)}
& \makecell{CommonVoice DE\\ASR (1k h)\\Conformer \cite{conformer}\\(110M)}
& \makecell{TED-LIUM 2\\ASR ($\sim$200h)\\Branchformer \cite{branchformer}\\(103M)}
\\
\midrule
No Aug
& 20.38 $\pm$ 0.06
& 16.37 $\pm$ 0.36
& 6.89 $\pm$ 0.03
& 11.41 $\pm$ 0.04
& 10.24 $\pm$ 0.10
\\
SpecAugment (LB) \cite{specaug}
& 19.79 $\pm$ 0.14
& 16.09 $\pm$ 0.44
& 5.98 $\pm$ 0.05
& \textbf{7.31 $\pm$ 0.07}
& \textbf{8.20 $\pm$ 0.02}
\\
SpecAugment (SM) \cite{specaug}
& \textbf{19.11 $\pm$ 0.06}
& \textbf{15.23 $\pm$ 0.41}
& \textbf{5.80 $\pm$ 0.08}
& 7.41 $\pm$ 0.08
& 8.26 $\pm$ 0.19
\\
\midrule
Wav2Aug top-6
& 19.60 $\pm$ 0.38
& 15.43 $\pm$ 0.30
& 6.08 $\pm$ 0.08
& 7.47 $\pm$ 0.10
& 8.35 $\pm$ 0.07
\\
Wav2Aug top-9
& 19.64 $\pm$ 0.46
& 15.54 $\pm$ 0.42
& 6.08 $\pm$ 0.08
& 7.48 $\pm$ 0.04
& 8.63 $\pm$ 0.08
\\
\bottomrule
\end{tabular}
\end{adjustbox}
\end{table*}

\subsection{ASR performance}
The following experiments assess whether Wav2Aug can maintain competitive speech recognition performance relative to a more established method, such as SpecAugment.
We evaluate speech and phoneme recognition on five benchmarks. These are TED-LIUM2 \cite{tedlium} with a BranchFormer model \cite{branchformer}, TIMIT \cite{timit} with a CRDNN model \cite{speechbrain_v1}, AISHELL-1 \cite{aishell} with a Transformer, CommonVoice Arabic \cite{commonvoice} with a fine-tuned Whisper model \cite{whisper}, and CommonVoice German \cite{commonvoice} with a Conformer \cite{conformer}. 
Importantly, we did not use any of these datasets during calibration. This separation lets us evaluate the proposed policy's generalization across unseen datasets, languages, architectures, and recognition settings.

Table~\ref{tab:asr} reports the recognition results across our five ASR benchmarks. Both Wav2Aug pools improve on non-augmented training, with top-6 reducing average relative error by 14.9\%. Compared with SpecAugment LB, top-6 improves on CommonVoice Arabic and TIMIT but trails on the other three benchmarks. On datasets measured by word error rate (WER), its largest gap to LB is 0.16 absolute percentage points. SpecAugment SM is stronger on Arabic, TIMIT, and AISHELL-1, with advantages over the top-6 of 0.49 WER, 0.20 phoneme error rate (PER), and 0.28 character error rate (CER) points, respectively.

These results suggest that Wav2Aug's broader augmentation policy does not improve ASR performance over SpecAugment's time and frequency masking, but it also does not substantially degrade it, a result we find interesting.

\begin{figure}[t]
  \centering
  \includegraphics[width=\columnwidth]{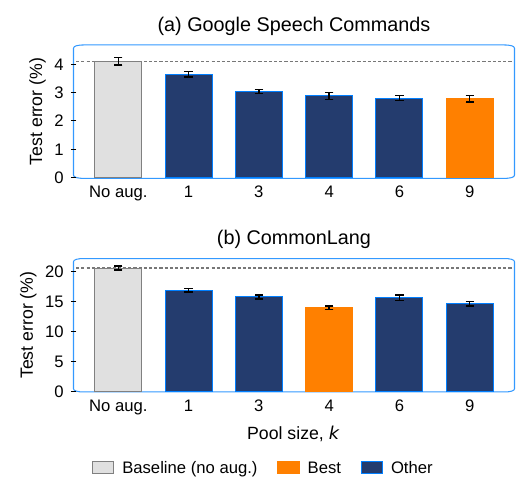}
  \caption{\normalfont Pool-size study on Google Speech Commands (top) and CommonLang (bottom), using the first $k$ ranked operations. Orange marks the lowest mean error in each panel. Bars show means and error bars show one sample standard deviation over five runs.}
  \label{fig:poolsize}
\end{figure}

\subsection{Transfer beyond ASR}
\label{sec:beyond_ASR}
We now evaluate Wav2Aug on tasks beyond ASR. Specifically, we consider five popular non-ASR speech tasks. These are keyword spotting on Google Speech Commands \cite{gsc} with X-Vectors \cite{xvector}, language identification on CommonLang \cite{commonlang} with ECAPA-TDNN \cite{ecapa}, emotion recognition on IEMOCAP \cite{iemocap} with wav2vec~2.0 \cite{wav2vec2}, spoken language understanding on SLURP \cite{slurp} with HuBERT \cite{hubert}, and voice activity detection on LibriParty VAD \cite{speechbrain_v1} with a CRDNN model. We train all models from scratch, except wav2vec~2.0 and HuBERT, which we fine-tune from pretrained checkpoints \cite{huggingface}.

\begin{table*}[t]
\centering
\caption{\normalfont Non-ASR test error rates (\%). Classification tasks use error rate, LibriParty uses detection error rate (DER), and SLURP uses WER. Mean $\pm$ standard deviation over five seeds; lower is better. Dashes indicate policies not compatible with the model architectures. Parameter counts are approximate.}
\label{tab:nonasr}
\small
\setlength{\tabcolsep}{4pt}
\renewcommand{\arraystretch}{1.12}
\begin{adjustbox}{max width=\textwidth}
\begin{tabular}{l *{5}{c}}
\toprule
Augmentation Policy
& \makecell{IEMOCAP (speech only)\\Emotion Rec. (12h)\\wav2vec2 \cite{wav2vec2}\\(94M)}
& \makecell{CommonLang\\LangID (45h) \\ECAPA-TDNN \cite{ecapa}\\(21M)}
& \makecell{Google Speech Cmd\\KWS (30h)\\X-Vectors \cite{xvector}\\(5M)}
& \makecell{LibriParty \\VAD (21h) \\CRDNN\\(0.11M)}
& \makecell{SLURP\\SLU (83h) \\HuBERT / Att-RNN \cite{hubert}\\(103M)}
\\
\midrule
No Aug
& 25.52 $\pm$ 1.82
& 20.55 $\pm$ 0.31
& 4.11 $\pm$ 0.14
& 9.19 $\pm$ 0.57
& 14.41 $\pm$ 0.23
\\
SpecAugment (latent) \cite{wav2vec2}
& 25.40 $\pm$ 2.58
& ---
& ---
& ---
& 13.96 $\pm$ 0.44
\\
SpecAugment (LB) \cite{specaug}
& ---
& 16.80 $\pm$ 0.67
& 5.06 $\pm$ 0.44
& 8.73 $\pm$ 0.52
& ---
\\
SpecAugment (SM) \cite{specaug}
& ---
& 17.08 $\pm$ 0.73
& 3.62 $\pm$ 0.09
& 8.44 $\pm$ 0.62
& ---
\\
\midrule
Wav2Aug top-6
& 23.50 $\pm$ 0.86
& 15.58 $\pm$ 0.47
& 2.81 $\pm$ 0.08
& \textbf{6.20 $\pm$ 0.17}
& \textbf{13.93 $\pm$ 0.19}
\\
Wav2Aug top-9
& \textbf{23.30 $\pm$ 1.48}
& \textbf{14.60 $\pm$ 0.39}
& \textbf{2.78 $\pm$ 0.11}
& 7.16 $\pm$ 0.19
& 14.24 $\pm$ 0.16
\\
\bottomrule
\end{tabular}
\end{adjustbox}
\end{table*}

\begin{table*}[t]
\centering
\small
\caption{\normalfont Multi-view results on CommonLang, AISHELL-1, and IEMOCAP. Mean $\pm$ std over 5 seeds. Avg. is the relative gain over one-view non-augmented training. Wav2Aug uses top-9 throughout.}
\label{tab:view_results}
\setlength{\tabcolsep}{8pt}
\renewcommand{\arraystretch}{1.15}
\begin{adjustbox}{max width=\textwidth}
\begin{tabular}{llcccc}
\toprule
\textbf{Batch Views} & \textbf{Policy} & \textbf{CommonLang} & \textbf{AISHELL-1} & \textbf{IEMOCAP} & \textbf{Avg.} \\
\midrule
$[C]$ & No aug. & 20.55 $\pm$ .31 & 6.89 $\pm$ .03 & 25.52 $\pm$ 1.82 & 0.0\% \\
$[A]$ & SpecAugment & 16.80 $\pm$ .67 & 5.98 $\pm$ .05 & 25.40 $\pm$ 2.58 & 10.6\% \\
$[A]$ & Wav2Aug & 14.60 $\pm$ .39 & 6.08 $\pm$ .08 & 23.30 $\pm$ 1.48 & 16.5\% \\
\midrule
$[C;A]$ & SpecAugment & 18.04 $\pm$ .45 & 5.99 $\pm$ .06 & 23.33 $\pm$ 1.69 & 11.3\% \\
$[C;A]$ & Wav2Aug & 14.62 $\pm$ .64 & 5.97 $\pm$ .06 & 22.56 $\pm$ 2.32 & 17.9\% \\
$[A;A]$ & SpecAugment & 16.62 $\pm$ .41 & 5.69 $\pm$ .04 & 23.08 $\pm$ 1.51 & 15.4\% \\
$[A;A]$ & Wav2Aug & \textbf{13.82 $\pm$ .36} & 5.84 $\pm$ .05 & 22.74 $\pm$ 1.81 & \textbf{19.6\%} \\
\midrule
$[C;A;A;A]$ & SpecAugment & 17.06 $\pm$ .48 & \textbf{5.66 $\pm$ .09} & 22.56 $\pm$ 1.21 & 15.5\% \\
$[C;A;A;A]$ & Wav2Aug & 15.14 $\pm$ .13 & 5.78 $\pm$ .08 & \textbf{21.58 $\pm$ 1.77} & 19.3\% \\
\bottomrule
\end{tabular}
\end{adjustbox}
\par\vspace{0.5ex}
{\raggedright\footnotesize C=non-augmented (clean), A=independently augmented. Metrics are classification error for CommonLang and IEMOCAP, and CER for AISHELL-1.\par}
\end{table*}

Table~\ref{tab:nonasr} reports results on the five non-ASR benchmarks. Both Wav2Aug pools improve over the no-augmentation condition on every task. With top-6, mean error is also lower than the strongest compatible SpecAugment baseline on all five tasks, yielding an average relative error reduction of 12.8\%. This average is computed by first taking the relative error reduction on each task and then averaging across tasks.

The gains depend on the task and on the baseline. On Google Speech Commands, SpecAugment LB degrades performance relative to no augmentation, whereas SM improves it. Top-6 further reduces error from 3.62\% with SM to 2.81\%. On LibriParty, top-6 reduces detection error from 8.44\% with SM to 6.20\%, a relative reduction of 26.5\%. It also improves over latent masking on IEMOCAP. On SLURP, however, the difference between top-6 and latent masking is only 0.03 WER points, much smaller than the reported standard deviations, so the result is better described as comparable performance.

The larger top-9 pool achieves the lowest mean errors on CommonLang, Google Speech Commands, and IEMOCAP, while top-6 performs better on LibriParty and SLURP. This demonstrates that adding operations does not uniformly improve performance, and neither pool is best on every task.

\subsection{Effect of the augmentation pool size}
\label{sec:poolsize}
To examine the effect of pool size more closely, we evaluate the first $k$ ranked operations for $k \in \{1,3,4,6,9\}$ on keyword spotting and language identification. The ranges remain fixed, and we sample two operations per batch whenever at least two are available. With a one-operation pool, we apply only noise addition.

Figure~\ref{fig:poolsize} shows that combining operations substantially improves performance over noise alone, but the best pool size differs by task. Google Speech Commands benefits from larger pools, with little difference between six and nine operations. CommonLang reaches its lowest mean error with four operations. Adding more does not improve performance monotonically. We conclude that no single pool size works best across all tasks, as certain operations may benefit specific tasks. Together with the ten benchmark results, these observations support top-6 as a practical default, rather than a universally optimal choice.

\subsection{Multi-view augmentation}
So far, all experiments have used the standard one-view setting, in which we augment each training utterance once before feeding it to the model. We now investigate whether exposing the model to multiple independently augmented versions of the same utterance within each training batch can further improve performance. Specifically, we concatenate a variable number of augmented views while keeping the number of unique utterances per batch fixed. For this study, Wav2Aug uses the top-9 policy throughout, SpecAugment uses LB on CommonLang and AISHELL-1, and latent masking on IEMOCAP. 

Table~\ref{tab:view_results} reports results for language identification on CommonLang, speech recognition on AISHELL-1, and emotion recognition on IEMOCAP. We evaluate several multi-view augmentation strategies, where C denotes the original (clean, non-augmented) input and A denotes an independently augmented version of the same utterance (using either SpecAugment or Wav2Aug).
In the last column, we report the average relative improvement over the non-augmented baseline for reference.

On average across the three tasks, multi-view training improves performance for both SpecAugment and Wav2Aug. This improvement comes at the cost of larger effective batch sizes, which increase both memory consumption and training time. However, the benefits of multi-view training quickly saturate, with four views offering little or no improvement over two views. On average, the best Wav2Aug configuration concatenates two independently augmented versions of each utterance within the same batch, without the original clean signal. This setting provides an average relative improvement of 19.6\% over the non-augmented baseline and an additional 3.1 percentage points over the single-view Wav2Aug setting. 

\section{Conclusion}
We introduced Wav2Aug, a simple drop-in waveform augmentation policy whose hyperparameter ranges are calibrated once and then reused unchanged across tasks and architectures. Across ten speech benchmarks, Wav2Aug consistently improves performance over the non-augmented baseline. The six-operation pool achieves a 12.8\% average relative error reduction over the strongest compatible SpecAugment baselines on non-ASR tasks, while remaining competitive with SpecAugment on ASR benchmarks. We also show that multi-view training provides modest additional gains, with two independently augmented views offering the best trade-off and an average relative improvement of 19.6\% over non-augmented training. Overall, Wav2Aug performs comparably to preset SpecAugment policies on ASR and, on average, better across the evaluated non-ASR tasks, without task-specific augmentation tuning.

\clearpage
\section*{Acknowledgements}
The first author thanks Firat Öncel for their guidance and thoughtful feedback throughout this work, and Francesco Bonzi for many helpful discussions. We acknowledge the Digital Research Alliance of Canada (DRAC) for providing computational resources. This work also used the LEONARDO supercomputer, owned by the EuroHPC Joint Undertaking and hosted by CINECA, Italy.

\IEEEtriggeratref{23}
\bibliographystyle{IEEEtran}
\bibliography{mybib}

@article{speechbrain_v1,
  author  = {M. Ravanelli and T. Parcollet and A. Moumen and others},
  title   = {Open-Source Conversational {AI} with {SpeechBrain} 1.0},
  journal = {Journal of Machine Learning Research},
  year    = {2024},
  volume  = {25},
  number  = {333},
  pages   = {1--11},
  url     = {http://jmlr.org/papers/v25/24-0991.html}
}

@inproceedings{specaug,
  author    = {D. S. Park and W. Chan and Y. Zhang and others},
  title     = {{SpecAugment}: A Simple Data Augmentation Method for Automatic Speech Recognition},
  booktitle = {Proc. of Interspeech},
  year      = {2019},
  pages     = {2613--2617},
  doi       = {10.21437/Interspeech.2019-2680}
}

@inproceedings{lu2020sentiment,
  author    = {Z. Lu and L. Cao and Y. Zhang and others},
  title     = {Speech Sentiment Analysis via Pre-Trained Features from End-to-End {ASR} Models},
  booktitle = {Proc. of ICASSP},
  year      = {2020},
  pages     = {7149--7153},
  doi       = {10.1109/ICASSP40776.2020.9052937}
}

@inproceedings{realm,
  title={{REAL-M: Towards speech separation on real mixtures}},
  author={Cem Subakan and Mirco Ravanelli and Samuele Cornell and François Grondin},
  booktitle={Proc. of ICASSP},
  year={2022},
}

@inproceedings{Ravanelli2016Realistic,
  title={Realistic Multi-Microphone Data Simulation for Distant Speech Recognition},
  author={Ravanelli, Mirco and Svaizer, Piergiorgio and Omologo, Maurizio},
  booktitle={Proc. of Interspeech},
  year={2016},
}

@inproceedings{le2021adapters,
  author    = {H. Le and J. Pino and C. Wang and others},
  title     = {Lightweight Adapter Tuning for Multilingual Speech Translation},
  booktitle = {Proc. of ACL-IJCNLP},
  year      = {2021},
  pages     = {817--824},
  doi       = {10.18653/v1/2021.acl-short.103}
}

@inproceedings{lee2022s2st,
  author    = {A. Lee and P.-J. Chen and C. Wang and others},
  title     = {Direct Speech-to-Speech Translation With Discrete Units},
  booktitle = {Proc. of ACL},
  year      = {2022},
  pages     = {3327--3339},
  doi       = {10.18653/v1/2022.acl-long.235}
}

@inproceedings{tang2021auxiliary,
  author    = {Y. Tang and J. Pino and X. Li and others},
  title     = {Improving Speech Translation by Understanding and Learning from the Auxiliary Text Translation Task},
  booktitle = {Proc. of ACL-IJCNLP},
  year      = {2021},
  pages     = {4252--4261},
  doi       = {10.18653/v1/2021.acl-long.328}
}

@inproceedings{tang2023hybrid,
  author    = {Y. Tang and A. Y. Sun and H. Inaguma and others},
  title     = {Hybrid Transducer and Attention based Encoder-Decoder Modeling for Speech-to-Text Tasks},
  booktitle = {Proc. of ACL},
  year      = {2023},
  pages     = {12441--12455},
  doi       = {10.18653/v1/2023.acl-long.695}
}

@inproceedings{ma2024simultaneous,
  author    = {Z. Ma and Q. Fang and S. Zhang and others},
  title     = {A Non-autoregressive Generation Framework for End-to-End Simultaneous Speech-to-Any Translation},
  booktitle = {Proc. of ACL},
  year      = {2024},
  pages     = {1557--1575},
  doi       = {10.18653/v1/2024.acl-long.85}
}

@inproceedings{rouditchenko2024flamingo,
  author    = {A. Rouditchenko and Y. Gong and S. Thomas and others},
  title     = {{Whisper-Flamingo}: Integrating Visual Features into {Whisper} for Audio-Visual Speech Recognition and Translation},
  booktitle = {Proc. of Interspeech},
  year      = {2024},
  pages     = {2420--2424},
  doi       = {10.21437/Interspeech.2024-322}
}

@inproceedings{whisper,
  author    = {A. Radford and J. W. Kim and T. Xu and others},
  title     = {Robust Speech Recognition via Large-Scale Weak Supervision},
  booktitle = {Proc. of ICML},
  year      = {2023},
  pages     = {28492--28518}
}

@article{qwen,
  author  = {Y. Chu and J. Xu and X. Zhou and others},
  title   = {{Qwen-Audio}: Advancing Universal Audio Understanding via Unified Large-Scale Audio-Language Models},
  journal = {arXiv preprint arXiv:2311.07919},
  year    = {2023}
}

@inproceedings{librispeech,
  author    = {V. Panayotov and G. Chen and D. Povey and others},
  title     = {{LibriSpeech}: An {ASR} Corpus Based on Public Domain Audio Books},
  booktitle = {Proc. of ICASSP},
  year      = {2015},
  pages     = {5206--5210},
  doi       = {10.1109/ICASSP.2015.7178964}
}

@article{dataaug,
  author  = {C. Shorten and T. M. Khoshgoftaar},
  title   = {A Survey on Image Data Augmentation for Deep Learning},
  journal = {Journal of Big Data},
  volume  = {6},
  number  = {1},
  pages   = {60},
  year    = {2019},
  doi     = {10.1186/s40537-019-0197-0},
  publisher = {Springer}
}

@inproceedings{wav2vec2,
  author    = {A. Baevski and Y. Zhou and A. Mohamed and others},
  title     = {{wav2vec} 2.0: A Framework for Self-Supervised Learning of Speech Representations},
  booktitle = {Proc. of NeurIPS},
  year      = {2020},
  pages     = {12449--12460}
}

@article{hubert,
  author  = {W.-N. Hsu and B. Bolte and Y.-H. H. Tsai and others},
  title   = {{HuBERT}: Self-Supervised Speech Representation Learning by Masked Prediction of Hidden Units},
  journal = {IEEE/ACM Transactions on Audio, Speech, and Language Processing},
  volume  = {29},
  pages   = {3451--3460},
  year    = {2021},
  doi     = {10.1109/TASLP.2021.3122291},
  publisher = {IEEE}
}

@inproceedings{ecapa,
  author    = {B. Desplanques and J. Thienpondt and K. Demuynck},
  title     = {{ECAPA-TDNN}: Emphasized Channel Attention, Propagation and Aggregation in {TDNN} Based Speaker Verification},
  booktitle = {Proc. of Interspeech},
  year      = {2020},
  pages     = {3830--3834},
  doi       = {10.21437/Interspeech.2020-2650}
}

@inproceedings{audrandaug,
  author  = {T. Kumar and M. Turab and A. Mileo and others},
  title   = {{AudRandAug}: Random Image Augmentations for Audio Classification},
  booktitle = {Proc. of IMVIP},
  year    = {2023},
  doi     = {10.5281/zenodo.8217347}
}

@inproceedings{speed_noise,
  author    = {T. Ko and V. Peddinti and D. Povey and others},
  title     = {Audio Augmentation for Speech Recognition},
  booktitle = {Proc. of Interspeech},
  year      = {2015},
  pages     = {3586--3589},
  doi       = {10.21437/Interspeech.2015-711}
}

@inproceedings{amplitude,
  author    = {Y. Zhang and Y. Zhang and T. Patel and others},
  title     = {Comparing Data Augmentation and Training Techniques to Reduce Bias Against Non-Native Accents in Hybrid Speech Recognition Systems},
  booktitle = {Proc. of S4SG},
  year      = {2022},
  pages     = {15--19},
  doi       = {10.21437/S4SG.2022-4}
}

@article{sign_flip,
  author  = {C. Rommel and T. Moreau and J. Paillard and others},
  title   = {{CADDA}: Class-Wise Automatic Differentiable Data Augmentation for {EEG} Signals},
  journal = {arXiv preprint arXiv:2106.13695},
  year    = {2021}
}

@inproceedings{babble,
  author    = {T. Prisyach and V. Mendelev and D. Ubskiy},
  title     = {Data Augmentation for Training of Noise Robust Acoustic Models},
  booktitle = {Proc. of AIST},
  year      = {2016},
  pages     = {17--25}
}

@inproceedings{gaugment,
  author    = {G. Wang and E. D. Cubuk and A. Rosenberg and others},
  title     = {{G-Augment}: Searching for the Meta-Structure of Data Augmentation Policies for {ASR}},
  booktitle = {Proc. of SLT},
  year      = {2023},
  pages     = {23--30}
}

@inproceedings{sapaugment,
  author    = {T.-Y. Hu and A. Shrivastava and J.-H. R. Chang and others},
  title     = {{SapAugment}: Learning a Sample Adaptive Policy for Data Augmentation},
  booktitle = {Proc. of ICASSP},
  year      = {2021},
  pages     = {4040--4044}
}

@inproceedings{commonvoice,
  author    = {R. Ardila and M. Branson and K. Davis and others},
  title     = {{Common Voice}: A Massively-Multilingual Speech Corpus},
  booktitle = {Proc. of LREC},
  year      = {2020},
  pages     = {4218--4222}
}

@inproceedings{voicebank,
  author    = {C. Veaux and J. Yamagishi and S. King},
  title     = {The {Voice Bank} Corpus: Design, Collection and Data Analysis of a Large Regional Accent Speech Database},
  booktitle = {Proc. of O-COCOSDA/CASLRE},
  year      = {2013},
  doi       = {10.1109/ICSDA.2013.6709856}
}

@inproceedings{xvector,
  author    = {D. Snyder and D. Garcia-Romero and G. Sell and others},
  title     = {{X-vectors}: Robust {DNN} Embeddings for Speaker Recognition},
  booktitle = {Proc. of ICASSP},
  year      = {2018},
  pages     = {5329--5333},
  doi       = {10.1109/ICASSP.2018.8461375}
}

@inproceedings{branchformer,
  author    = {Y. Peng and S. Dalmia and I. Lane and others},
  title     = {{Branchformer}: Parallel {MLP}-Attention Architectures to Capture Local and Global Context for Speech Recognition and Understanding},
  booktitle = {Proc. of ICML},
  year      = {2022},
  pages     = {17627--17643}
}

@inproceedings{conformer,
  author    = {A. Gulati and J. Qin and C.-C. Chiu and others},
  title     = {{Conformer}: Convolution-Augmented Transformer for Speech Recognition},
  booktitle = {Proc. of Interspeech},
  year      = {2020},
  pages     = {5036--5040},
  doi       = {10.21437/Interspeech.2020-3015}
}

@inproceedings{randaug,
  author    = {E. D. Cubuk and B. Zoph and J. Shlens and others},
  title     = {{RandAugment}: Practical Automated Data Augmentation with a Reduced Search Space},
  booktitle = {Proc. of CVPR Workshops},
  year      = {2020},
  pages     = {3008--3017},
  doi       = {10.1109/CVPRW50498.2020.00359}
}

@inproceedings{pointsource,
  author    = {T. Ko and V. Peddinti and D. Povey and others},
  title     = {A Study on Data Augmentation of Reverberant Speech for Robust Speech Recognition},
  booktitle = {Proc. of ICASSP},
  year      = {2017}
}

@inproceedings{tedlium,
  author    = {A. Rousseau and P. Deli{\`e}glise and Y. Esteve and others},
  title     = {Enhancing the {TED-LIUM} Corpus with Selected Data for Language Modeling and More {TED} Talks},
  booktitle = {Proc. of LREC},
  year      = {2014},
  pages     = {3935--3939}
}

@techreport{timit,
  author  = {J. S. Garofolo and L. F. Lamel and W. M. Fisher and others},
  title   = {{DARPA TIMIT} Acoustic-Phonetic Continuous Speech Corpus {CD-ROM}, {NIST} Speech Disc 1-1.1},
  institution = {National Institute of Standards and Technology},
  number  = {NIST IR 4930},
  year    = {1993},
  doi     = {10.6028/NIST.IR.4930}
}

@inproceedings{aishell,
  author    = {H. Bu and J. Du and X. Na and others},
  title     = {{AISHELL-1}: An Open-Source Mandarin Speech Corpus and a Speech Recognition Baseline},
  booktitle = {Proc. of O-COCOSDA},
  year      = {2017},
  pages     = {1--5},
  doi       = {10.1109/ICSDA.2017.8384449}
}

@article{gsc,
  author  = {P. Warden},
  title   = {Speech Commands: A Dataset for Limited-Vocabulary Speech Recognition},
  journal = {arXiv preprint arXiv:1804.03209},
  year    = {2018}
}

@misc{commonlang,
  author    = {G. Sinisetty and P. Ruban and O. Dymov and others},
  title     = {{CommonLanguage}},
  month     = jun,
  year      = {2021},
  publisher = {Zenodo},
  version   = {0.1},
  doi       = {10.5281/zenodo.5036977},
  url       = {https://doi.org/10.5281/zenodo.5036977}
}

@article{iemocap,
  author  = {C. Busso and M. Bulut and C.-C. Lee and others},
  title   = {{IEMOCAP}: Interactive Emotional Dyadic Motion Capture Database},
  journal = {Language Resources and Evaluation},
  volume  = {42},
  number  = {4},
  pages   = {335--359},
  year    = {2008},
  doi     = {10.1007/s10579-008-9076-6},
  publisher = {Springer}
}

@inproceedings{slurp,
  author    = {E. Bastianelli and A. Vanzo and P. Swietojanski and others},
  title     = {{SLURP}: A Spoken Language Understanding Resource Package},
  booktitle = {Proc. of EMNLP},
  year      = {2020},
  pages     = {7252--7262},
  doi       = {10.18653/v1/2020.emnlp-main.588}
}

@inproceedings{huggingface,
  author    = {T. Wolf and L. Debut and V. Sanh and others},
  title     = {Transformers: State-of-the-Art Natural Language Processing},
  booktitle = {Proc. of EMNLP: System Demonstrations},
  year      = {2020},
  pages     = {38--45},
  doi       = {10.18653/v1/2020.emnlp-demos.6}
}

@inproceedings{contaminated_speech_mirco,
  author    = {M. Ravanelli and M. Omologo},
  title     = {Contaminated Speech Training Methods for Robust {DNN-HMM} Distant Speech Recognition},
  booktitle = {Proc. of Interspeech},
  year      = {2015},
  pages     = {756--760},
  doi       = {10.21437/Interspeech.2015-251}
}

@inproceedings{impulse_mirco,
  author    = {M. Ravanelli and M. Omologo},
  title     = {On the Selection of the Impulse Responses for Distant-Speech Recognition Based on Contaminated Speech Training},
  booktitle = {Proc. of Interspeech},
  year      = {2014},
  pages     = {1028--1032},
  doi       = {10.21437/Interspeech.2014-268}
}

@inproceedings{autoaugment,
  author    = {E. D. Cubuk and B. Zoph and D. Man{\'e} and others},
  title     = {{AutoAugment}: Learning Augmentation Strategies From Data},
  booktitle = {Proc. of CVPR},
  year      = {2019},
  pages     = {113--123},
  doi       = {10.1109/CVPR.2019.00020}
}

@article{wavlm,
  title={{WavLM}: Large-Scale Self-Supervised Pre-Training for Full Stack Speech Processing},
  author={Chen, Sanyuan and Wang, Chengyi and Chen, Zhengyang and Wu, Yu and Liu, Shujie and Chen, Zhuo and Li, Jinyu and Kanda, Naoyuki and Yoshioka, Takuya and Xiao, Xiong and others},
  journal={IEEE Journal of Selected Topics in Signal Processing},
  volume={16},
  number={6},
  pages={1505--1518},
  year={2022},
  doi={10.1109/JSTSP.2022.3188113},
  publisher={IEEE}
}

@inproceedings{hoffer2020augment,
  title     = {Augment Your Batch: Improving Generalization Through Instance Repetition},
  author    = {Hoffer, Elad and Ben-Nun, Tal and Hubara, Itay and Giladi, Niv and Hoefler, Torsten and Soudry, Daniel},
  booktitle = {Proc. of CVPR},
  pages     = {8129--8138},
  year      = {2020},
  doi       = {10.1109/CVPR42600.2020.00815}
}

@article{fort2021drawing,
  title   = {Drawing Multiple Augmentation Samples Per Image During Training Efficiently Decreases Test Error},
  author  = {Fort, Stanislav and Brock, Andrew and Pascanu, Razvan and De, Soham and Smith, Samuel L.},
  journal = {arXiv preprint arXiv:2105.13343},
  year    = {2021}
}

@article{kivaisi2024minibatch,
  title   = {Improved Mini-Batch Multiple Augmentation for Low-Resource Spoken Word Recognition},
  author  = {Kivaisi, Alexander Rogath and Zhao, Qingjie},
  journal = {Expert Systems with Applications},
  volume  = {252},
  pages   = {124157},
  year    = {2024},
  doi     = {10.1016/j.eswa.2024.124157}
}

@inproceedings{singh2024roar,
  title     = {{ROAR}: Reinforcing Original to Augmented Data Ratio Dynamics for {wav2vec} 2.0 Based {ASR}},
  author    = {Singh, Vishwanath Pratap and Malato, Federico and Hautam{\"a}ki, Ville and Sahidullah, Md. and Kinnunen, Tomi},
  booktitle = {Proc. of Interspeech},
  pages     = {2885--2889},
  year      = {2024},
  doi       = {10.21437/Interspeech.2024-1403}
}

\end{document}